\documentclass[12pt]{article}
\usepackage{amsfonts}

\usepackage{graphicx}
\usepackage{amsmath}

\begin{document}

\title{Generalized Bogoliubov Transformation for Confined Fields: Applications in
Casimir Effect}
\author{J. C. da Silva$^{a,b}$, F. C. Khanna$^{c,d},$ A. Matos Neto$^{a}$ $\,\,$ ,
\and  A. E. Santana$^{a,c}$. \\
${}^a$ Instituto de F\'\i sica, Universidade Federal da Bahia,\\
Campus de Ondina, 40210-340, Salvador, Bahia, Brasil.\\
$^b$ Centro Federal de Educa\c c\~ao Tecnol\'ogica da Bahia\\
Rua Em\'{i}dio Santos, 40000-900, Salvador, Bahia, Brasil\\
$^{c}$Physics Department, Theoretical Physics Institute,\\
University of Alberta, Edmonton, Alberta T6G 2J1 Canada\\
$^{d}$TRIUMF, 4004, Westbrook mall, Vancouver, British \\
Columbia V6T 2A3, Canada\\
}
\maketitle

\begin{abstract}
The Bogoliubov transformation in thermofield dynamics, an operator formalism
for the finite-temperature quantum-field theory, is generalized to describe
a field in arbitrary confined regions of space and time. Starting with the
scalar field, the approach is extended to the electromagnetic field and the
energy-momentum tensor is written via the Bogoliubov transformation. In this
context, the Casimir effect is calculated for zero and non-zero temperature,
and therefore it can be considered as a vacuum condensation effect of the
electromagnetic field. This aspect opens an interesting perspective for
using this procedure as an effective scheme for calculations in the studies
of confined fields, including the interacting fields.
\end{abstract}

\section{Introduction}

A way to treat the effect of temperature in quantum field theory is, for
instance, through the  Matsubara formalism\cite{mat1,le1}, which is based on
a formal substitution of time, say $t$, by a complex time, say $i\tau .$ In
this imaginary time scheme, the temperature emerges as a consequence of a
compactification of the field in a finite interval in time axis, $0<\tau
<\beta ,$ where $\beta $ is the inverse of temperature (we take the
Boltzmann constant as $k_{B}=1;$ thus $\beta =1/k_{B}T=1/T)$. This
compactification effect of time coordinates has been generalized and
associated with space confinement, for instance, of the electromagnetic
field through the notion of image method for a Green
 function\cite{brown,plu1}. Besides that, in the realm of euclidean
theories, a generalization of the Matsubara formalism has been carried out
to take into account spatial confinement of the scalar field in the $\lambda
\phi ^{4}$ approach using the Epstein-Hurwitz zeta-functions \cite
{jorge1,jorge2,eliz1}.

On the other hand, as an alternative finite-temperature quantum field
theory, Takahashi and Umezawa introduced the so-called thermofield dynamics
(TFD)
approach\cite{ume2,ume4,rev2,ari2,ume1,kha2,kha3,ade1,kha1,
kob1},  in order to handle finite temperature with a real
time operator formalism\cite {kob1,leb1}. TFD is based on
two elements. The first one is a doubling in the Fock space
${\cal H}$ of the original field system, giving rise to an
expanded Fock space denoted by ${\cal H}_{T}={\cal
H}\otimes \widetilde{ {\cal H}}$. This doubling, in terms
of mappings in ${\cal H}_{T}$, is defined by what is called
tilde (or dual) conjugation rules, associating each
operator in ${\cal H}$, say $a$, to two operators in ${\cal
H}_{T}$, say $A=a\otimes 1$ and $\widetilde{A}=1\otimes a,$
such that the physical variables are described by the
non-tilde operators. The next basic ingredient of TFD is a
Bogoliubov transformation, introducing a rotation in the
tilde and non-tilde variables, such that the temperature
effect emerges from a condensed state.

When TFD is compared with the imaginary-time (Matsubara) formalism, the
Bogoliubov transformation works by confining the field in a restricted
region of the time axis via the notion of a condensate. Here our main goal
is to develop a generalization of TFD to extend the compactification of an
arbitrary field for regions in space through the notion of condensate. This
possibility finds support and should be useful in different contexts, in
particular when associated with the vacuum properties of the electromagnetic
field via the Casimir effect.

The Casimir effect arises from the fluctuation of the vacuum state of fields
defined in space-time manifolds with non-trivial topologies\cite
{cas1,milon,mostep1,milton1,car1,sva1,car2,far1,far2}. Nowadays there is a
prominent interest in Casimir effect, as a consequence of the fact that in
1997 the Casimir force was measured with the precision of a few percent\cite
{lam2,roy,lam1,bordag,mostep2}. This result points in particular to a
practical use of the Casimir force in nano- and micro-technologies\cite
{lam1,mostep2}. Nevertheless, the finite-temperature corrections in the
Casimir effect are one aspect which still demand more theoretical
development and need greater experimental accuracy. Indeed, for example,
only recently the implications of temperature have been analysed in the
context of the classical limit of the Casimir effect\cite
{mann1,mann11,mann3}, although, as it was emphazised by Mehra long ago
\cite{mehra}, that the temperature effect is significant for plates
separated by a distance of the order of micrometers.

Brown and Maclay\cite{brown}  treated temperature in the
Casimir effect in a full finite-temperature quantum-field theory point of
view. In their approach the energy-momentum tensor for the electromagnetic
field is written in the context of the imaginary time formalism and the
image-method procedure is used to calculate the propagator between two
parallel plates. As a result the temperature effect emerges in the time axis
by using the set of images for the propagator. In this calculation, all
infinite images are summed up. A set of them (also infinite in number and
called odd-images), associated with an attempt to make the formalism
somewhat covariant, gives zero contribution by a resummation of the infinite
series but only at the end of calculation. This remains as a difficulty,
elucidated in the present formalism, since the algebraic manipulations are
numerous and there is no available criterion to specify which set of images
gives the proper contribution in any specific case\cite{brown,plu1}. Here we
overcome those difficulties by  using a generalization of TFD via an
analytic continuation of the Bogoliubov transformations. In this case, we
are not concerned with images, such that the sum over odd-images is no
longer needed. Furthermore, the calculations are carried out in a natural
covariant way. These aspects represent an ease of calculations, following a
rigorous procedure, with an unexpected elucidation of the whole process in a
confined space, the Casimir force, derived via the Bogoliubov
transformation, can be thought of as a vaccum condensation effect of the
electromagnetic field.

We have also applied the method to the
situation of one plate made of metal and the other of
permeable material (the Casimir-Boyer model)\cite{boyer2}.
The interest for this systems lies in the fact that we have
a repulsive Casimir force, and this type of  force has not
been satisfactorily studied, although it is a fundamental
ingredient to use the Casimir effect associated with
technological nanodevices\cite{kenne}. Furthermore, the
Casimir-Boyer model has  been only recently addressed in
more detail in the literature, in particular in the study
of temperature effects \cite{boyer2,boy2,boy3}. Regarding
this sytem, the new results  derived here include an
explicit expression for the energy-momentum tensor and an
analysis of thermodynamics functions such as  Helmholtz
free energy  and entropy. The results thus obtained  open
an interesting perspective to use this procedure as an
effective scheme for calculations in the studies of
different confined fields, including non-abelian gauge
fields like QCD. In other words, this generalized TFD
formalism  identifies the Casimir effect as giving a 
direct and clear picture of the vacuum as it displays its
properties for differents fields including the non-abelian
field.

In order to proceed with, in Section 2 the notation describing the basic
elements of TFD is set forth, and the generalized Bogoliubov transformation
is introduced for the scalar and the electromagnetic field. The subsequent
sections are dedicated to applications. In Section 3, a thermal
stress-energy tensor for the electromagnetic field is derived, and the
results are compared with those from the imaginary-time formalism. In
Section 4, the Casimir effect at zero and non-zero temperature are derived.
In this case we consider the field constrained between two
parallel plates (both made of either conducting or
permeable material). The mixed situation, the Casimir-Boyer
model\cite{boyer2,boy2,boy3}, in which there is one
conduting and one permeable plate, is discussed in Section
5. Our final remarks and conclusions are presented in
section 6.

\section{TFD and generalized Bogoliubov transformation}

Thermofield dynamics is introduced by assuming that the set of operators in
a field theory can be given in the form ${\cal L}_{T}=\{A,B,C,...,\widetilde{
A},\widetilde{B},\widetilde{C}...\}$, defined in the
Hilbert space ${\cal H} _{T}={\cal H}\otimes
\widetilde{{\cal H}}$with elements $|\Phi \rangle =|\phi
,\widetilde{\phi }\rangle $. The action of generic
operators $A$ and $ \widetilde{A}$ on $|\Phi \rangle $ is
specified by \begin{eqnarray} A|\Phi \rangle  &\equiv
&a\otimes 1(|\phi \rangle \otimes \langle \phi |)=(a|\phi
\rangle )\otimes \langle \phi |,  \label{at1} \\
\widetilde{A}|\Phi \rangle  &=&1\otimes a(|\phi \rangle
\otimes \langle \phi |)=|\phi \rangle \otimes \langle \phi
|a^{\dagger },  \label{at2} \end{eqnarray} where the
operator $a$ is defined in the usual Hilbert space, ${\cal
H}$, with $|\phi \rangle \in {\cal H}$ (We follow the usual
notation, which is introduced via the identification: $A=a$
and $\widetilde{A}=\widetilde{a}$). The tilde (or dual)
conjugation rules are then a mapping \ $\widetilde{} 
:A\rightarrow \widetilde{A},$ specified by the relations
(which can be derived from general algebraic
properties\cite{kha5,mar1}), \begin{eqnarray}
(A_{i}A_{j})\widetilde{}
&=&\widetilde{A}_{i}\widetilde{A}_{j}, \label{c1til2} \\
(cA_{i}+A_{j})\widetilde{} &=&c^{\ast
}\widetilde{A}_{i}+\widetilde{A}_{j}, \label{c1til3} \\
(A_{i}^{\dagger })\widetilde{}
&=&(\widetilde{A}_{i})^{\dagger }, \label{c1til4} \\
(\widetilde{A}_{i})\widetilde{} &=&A_{i}  \label{c1til5} \\
\lbrack \widetilde{A}_{i},A_{j}] &=&0.  \label{c1till6}
\end{eqnarray}

The physical variables are described by the non-tilde operators and thermal
variables are introduced by a Bogoliubov transformation defined by the
following procedure. For an arbitrary bosonic operator $A$ we define 
\begin{equation}
(A^{a})=\left(
\begin{array}{c}
A\  \\
-\widetilde{A}^{\dagger }\
\end{array}
\right) ,\,\,\,(\,A^{a\dagger })=\left( A^{\dagger }\ \,\,,\,\,\widetilde{A}
\ \right) .\,\,  \label{doubb1}
\end{equation}
The Bogoliubov transformation is defined as a $2\times 2$ matrix, 
\begin{equation}
{\cal B}=\left( 
\begin{array}{cc}
\,\,\,\,\,u\,\,(\alpha ) & -v(\alpha ) \\ 
-v(\alpha ) & \,\,u(\alpha )
\end{array}
\right) ,  \label{bog111}
\end{equation}
where $u^{2}(\alpha )-v^{2}(\alpha )=1,$ with $\alpha $ being a parameter
specifying the rotation between tilde and non-tide variables. For instance,
for the case of the creation $(a^{\dagger },\widetilde{a}^{\dagger })$ and
destruction $(a,\widetilde{a})$ boson operators, we have the extended
(doubled) algebra 
\begin{eqnarray}
\lbrack a,a^{\dagger }] &=&[\widetilde{a},\widetilde{a}^{\dagger }]=1,
\label{bos1} \\
\lbrack a^{\dagger },\widetilde{a}] &=&[a^{\dagger },\widetilde{a}^{\dagger
}]=[a,\widetilde{a}]=[a,\widetilde{a}^{\dagger }]=0.  \label{bos2}
\end{eqnarray}
Then the algebraic rules for the thermal bosonic operators are written as $
[a^{a}(\alpha ),a^{b\dagger }(\alpha )]=\delta ^{ab};$ $\,\,a,b=1,2,$ such
that $a^{a}=({\cal B}^{-1})^{ab}a^{b}(\alpha
)\,\,\,$and$\,\,\,\,\,\,\, \,a^{a\dagger }=a^{b}(\alpha 
)\,{\cal B}^{ba}\,.$ Writing explicitly, we have
\begin{eqnarray}
a &=&u(\alpha )\,a(\alpha )+v(\alpha )\,\widetilde{a}^{\dagger }(\alpha ) \\
\widetilde{a} &=&u(\alpha )\,\widetilde{a}(\alpha )+v(\alpha )\,a^{\dagger
}(\alpha ). \\
a^{\dagger } &=&u(\alpha )a^{\dagger }(\alpha )+v(\alpha
)\widetilde{a}(\alpha ), \\
\widetilde{a}^{\dagger } &=&u(\alpha )\widetilde{a}^{\dagger }(\alpha
)+v(\alpha )a(\alpha ).
\end{eqnarray}

The thermal average is given by taking the vacuum avarage
$|0,\widetilde{0} \rangle $ of a thermal non-tilde
variables. For instance for the particular case of $\ $\
the bosonic number operator, $n=a^{\dagger }a$, the thermal
distribution is given by \begin{equation}
n(\beta )=\langle 0,\widetilde{0}\rangle |a^{\dagger
}(\alpha )a(\alpha )|0, \widetilde{0}\rangle \equiv
\frac{1}{e^{\beta \epsilon }-1}  =\sum_{n=1}^{\infty
}e^{-\beta \varepsilon n}.  \label{distr} \end{equation}
This result is readily derived if the parameter $\alpha $ is taken to be the
temperature, $\alpha =\beta ,$ such that 
\begin{eqnarray}
u(\beta ) &=&\frac{1}{[1-e^{-\beta \epsilon }]^{1/2}},  \label{cos1} \\
v(\beta ) &=&\frac{1}{[e^{\beta \epsilon }-1]^{1/2}}.  \label{cos2}
\end{eqnarray}
However, obeserve that $\alpha =\beta $ is a  particular
choice, and other possibilities for $\alpha $ can be
considered as it will be seen in the next sections.

The same scheme is generalized for a quantum field. In the case of the free
scalar field, since we have equation of motion for the tilde and non-tilde
variables, the $\alpha -$dependent Klein-Gordon field theory is given by the
Lagrangian 
\[
\widehat{L}=\frac{1}{2}\partial _{\mu }\phi (x;\alpha )\partial ^{\mu }\phi
(x;\alpha )-\frac{m^{2}}{2}\phi (x;\alpha )^{2}-\frac{1}{2}\partial _{\mu }
\widetilde{\phi }(x;\alpha )\partial ^{\mu }\widetilde{\phi }(x;\alpha )+
\frac{m^{2}}{2}\widetilde{\phi }(x;\alpha )^{2},
\]
where the metric $g^{\mu \lambda }$ is such that $diag(g^{\mu \lambda
})=(1,-1,-1,-1).$ This Lagrangian gives rise to the equations of motions 
\[
(\partial _{\mu }\partial ^{\mu }+m^{2})\phi (x;\alpha
)=0,\;\;\;\;and\;\;\;\;(\partial _{\mu }\partial ^{\mu }+m^{2})\widetilde{
\phi }(x;\alpha )=0.
\]
Therefore, in TFD the Lagrangian can be written as
$\widehat{L}=L-\widetilde{L  }$ and in consequence the
Hamiltonian is $\widehat{H}=H-\widetilde{H}$ (this is a
general result which can be used for every field). The
two-point Green function for the $\alpha -$scalar field is
defined, then, by
\begin{eqnarray} G(x-x^{\prime };\alpha
)^{(ab)} &=&\langle 0,\widetilde{0}|T[\phi (x;\alpha
)^{a}\phi (x^{\prime };\alpha )^{b}|0,\widetilde{0}\rangle
\nonumber \\
&=&\frac{1}{(2\pi )^{4}}\int d^{4}k\,G(k;\alpha
)^{ab}e^{ik(x-x^{\prime })}\label{eq24},
\end{eqnarray} where
\[
G(k;\alpha )^{(ab)}={\cal B}^{-1}(k_{o};\alpha
)G_{o}(k)^{ab}{\cal B} (k_{o};\alpha ),
\]
with 
\begin{equation}
{\cal B}(k,\alpha )=\left( 
\begin{array}{cc}
\,\,\,\,\,u(k,\alpha ) & -v(k,\alpha ) \\ 
-v(k,\alpha ) & \,\,u(k,\alpha )
\end{array}
\right) ,  \label{and1}
\end{equation}
and

\begin{equation}
(G_{o}(k)^{ab})=\left( 
\begin{array}{cc}
G_{0}(k) & 0 \\ 
0 & \widetilde{G}_{0}(k)
\end{array}
\right) =\left( 
\begin{array}{cc}
\frac{1}{k^{2}-m^{2}+i\epsilon } & 0 \\
0 & \frac{-1}{k^{2}-m^{2}-i\epsilon }
\end{array}
\right) .  \label{adefour1}
\end{equation}
Then we have 
\begin{eqnarray*}
G(k;\alpha )^{(ab)} &=&{\cal B}(k,\alpha )G(k)^{(ab)}{\cal B}(k,\alpha ) \\
&=&\left(
\begin{array}{cc}
\,\,G(k;\alpha )^{11}\,\,\, & G(k;\alpha )^{12} \\
G(k;\alpha )^{21} & G(k;\alpha )^{22}
\end{array}
\right) ,
\end{eqnarray*}
with
\begin{eqnarray*}
\,\,G(k;\alpha )^{11}\, &=&G_{0}(k)+v^{2}(k,\alpha )[G_{0}(k)+\widetilde{G}
_{0}(k)], \\
\,G(k;\alpha )^{12}\, &=&G(k;\alpha )^{21}\,=v(k,\alpha )[1+v^{2}(k,\alpha
)]^{1/2}[G_{0}(k)+\widetilde{G}_{0}(k)], \\
G(k;\alpha )^{22}\, &=&\widetilde{G}_{0}(k)+v^{2}(k,\alpha )[G_{0}(k)+
\widetilde{G}_{0}(k)].
\end{eqnarray*}
The physical information is given by $G(k;\alpha )^{11}\,.$

Using the definition of ${\cal B}(k_{o},\alpha )$ given in Eq.(\ref{and1}),
with $n(k,\alpha =\beta )=v(k,\alpha =\beta )^{2}=1/[e^{\beta k_{o}}-1],$and
$u(k,\alpha =\beta )^{2}=v(k,\alpha =\beta )^{2}+1=1/[1-e^{-\beta k_{o}}]$,
the components of $G(k;\beta )^{ab}$ read 
\begin{eqnarray*}
G(k;\beta )^{(11)} &=&\frac{1}{k^{2}-m^{2}+i\epsilon }-2\pi i\,n(k_{o};\beta
)\,\delta (k^{2}-m^{2}), \\
G(k;\beta )^{(22)} &=&\frac{-1}{k^{2}-m^{2}-i\epsilon }-2\pi
i\,n(k_{o};\beta )\,\delta (k^{2}-m^{2}), \\
G(k;\beta )^{(12)} &=&G(k;\beta )^{21}=-2\pi i\,[n(k_{o};\beta
)+n(k_{o};\beta )^{2}]^{1/2}\,\delta (k^{2}-m^{2}),
\end{eqnarray*}
where $n(k_{o};\beta )=$ $u(k,\alpha =\beta )^{2}$ is the boson distribution
function. Notice that the physical propagator $G(k;\beta )^{11}$ is a well
known result derived with the Matsubara method.

The components of \[
G(x-x^{\prime })^{(ab)}=\langle 0,\widetilde{0}|T[\phi (x)^{a}\phi
(x^{\prime })^{b}|0,\widetilde{0}\rangle
\]
for a massless bosonic field can be explicitly written, by
using Eq.(\ref{adefour1}), as \begin{eqnarray}
G_{0}^{(11)}(x-x^{\prime }) &\equiv &G_{0}(x-x^{\prime })=-\frac{i}{(2\pi
)^{2}}\frac{1}{(x-x^{\prime })^{2}+i\eta },  \label{gree2} \\
G_{0}^{(22)}(x-x^{\prime }) &\equiv &\widetilde{G}_{0}(x-x^{\prime })=\frac{i
}{(2\pi )^{2}}\frac{1}{(x-x^{\prime })^{2}-i\eta },  \label{gree3}
\end{eqnarray}
and $G_{0}^{(12)}(x-x^{\prime })=G_{0}^{(21)}(x-x^{\prime
})=0.$  These results will be useful in the following
development.

Now we consider the\ case of electromagnetic field. Following the tilde
conjugation rules, the doubled operator describing the energy-momentum
tensor of the electromagnetic field is, then , given by 
\begin{equation}
T^{\mu \lambda (ab)}=-F^{\mu \alpha (ab)}F_{\,\,\,\,\alpha }^{\lambda (ab)}+
\frac{1}{4}g^{\mu \lambda }F_{\,\beta \alpha }^{(ab)}F^{\alpha \beta (ab)},
\label{ten1}
\end{equation}
where the non-tensorial indices $a,b=1,2$ \ are defined according to the
doubled notation given in Eqs. (\ref{doubb1}), 
\[
F_{\mu \nu }^{(ab)}=\partial _{\mu }A_{\nu }^{a}-\partial _{\nu }A_{\mu
}^{b}. 
\]
The doubled free-photon propagator is thus given by 
\begin{eqnarray}
iD_{\alpha \beta }^{(ab)}(x-x^{\prime }) &=&\langle
0,\widetilde{0} |T[A_{\alpha }^{a}(x)A_{\beta
}^{b}(x)]|0,\widetilde{0}\rangle  \nonumber \\ &=&g_{\alpha
\beta }G_{0}^{(ab)}(x-x^{\prime }),  \label{gree1}
\end{eqnarray}
where the non-zero components of
$G_{0}^{(ab)}(x-x^{\prime })$ are given in
Eqs.(\ref{gree2}) and (\ref{gree3})

The vacuum average of the energy-momentum tensor reads 
\begin{eqnarray*}
\langle 0,\widetilde{0}|T^{\mu \nu (ab)}|0,\widetilde{0}\rangle 
&=&-i\{\Gamma ^{\mu \nu }(x,x^{\prime })G_{0}^{(ab)}(x-x^{\prime }) \\
&&+2(\eta ^{\mu }\eta ^{\nu }-\frac{1}{4}g^{\mu \nu })\delta (x-x^{\prime
})\delta ^{ab}\}|_{x\rightarrow x^{\prime }},
\end{eqnarray*}
where $\eta ^{\mu }=(1,0,0,0)$ and 
\[
\Gamma ^{\mu \nu }(x,x^{\prime })=2(\partial ^{\mu
}\partial ^{\prime \nu }- \frac{1}{4}g^{\mu \nu }\partial
^{\rho }\partial _{\rho }^{\prime }). \]

Inspired by the usual Casimir prescription, at this point we introduce the
tensor $T^{\mu \lambda (ab)}(\alpha )$ by 
\begin{equation}
T_{cas}^{\mu \lambda (ab)}(x;\alpha )=\langle 0,\widetilde{0}|T^{\mu \lambda
(ab)}(x;\alpha )|0,\widetilde{0}\rangle -\langle 0,\widetilde{0}|T^{\mu
\lambda (ab)}(x)|0,\widetilde{0}\rangle ,  \label{casad1}
\end{equation}
where 
\begin{eqnarray*}
\langle 0,\widetilde{0}|T^{\mu \lambda (ab)}(x;\alpha
)|0,\widetilde{0} \rangle &=&-i\{\Gamma ^{\mu \lambda
}(x,x^{\prime })G^{(ab)}(x-x^{\prime };\alpha ) \\
&&+2(\eta ^{\mu }\eta ^{\lambda}-\frac{1}{4}g^{\mu \lambda})\delta (x-x^{\prime
})\delta ^{ab}\}|_{x\rightarrow x^{\prime }}.
\end{eqnarray*}
As a consequence 
\begin{equation}
T_{cas}^{\mu \lambda (ab)}(x;\alpha )=-i\{\Gamma ^{\mu
\lambda}(x,x^{\prime }) \overline{G}^{(ab)}(x-x^{\prime
};\alpha )\}|_{x\rightarrow x^{\prime }}, \label{emtenso1}
\end{equation}
where 
\[
\overline{G}^{(ab)}(x-x^{\prime };\alpha )=\frac{1}{(2\pi )^{4}}\int
d^{4}k\,e^{ik(x-x^{\prime })}\overline{G}^{(ab)}(k;\alpha ), 
\]
such that 
\begin{eqnarray*}
\overline{G}^{(11)}(k;\alpha ) &=&\overline{G}^{(22)}(k;\alpha
)=v^{2}(k,\alpha )[G_{0}(k)+\widetilde{G}_{0}(k)], \\
\overline{G}^{(12)}(k;\alpha ) &=&\overline{G}^{(21)}(k;\alpha )=v(k,\alpha
)[1+v^{2}(k,\alpha )]^{1/2}[G_{0}(k)+\widetilde{G}_{0}(k)].
\end{eqnarray*}

Let us write a general form for the Bogoliubov transformation by assuming
the following analytical continuation for $v^{2}(k,\alpha )$, given
originally in Eq.(\ref{distr}), 
\begin{equation}
v^{2}(k,\alpha )\equiv \sum_{l=1}^{\infty }e^{-i\alpha _{l}\cdot k},
\label{genbog}
\end{equation}
where the notation is 
\[
\sum_{l=1}^{\infty }e^{-i\alpha _{l}\cdot
k}=\sum_{l_{0},l_{1},l_{2},l_{3}=1}^{\infty }\exp [-i(\alpha
_{0}l_{0}k_{0}+\alpha _{1}l_{1}k_{1}+\alpha _{2}l_{2}k_{2}+\alpha
_{3}l_{3}k_{3})], 
\]
with $\alpha =(\alpha _{0},\alpha _{1},\alpha _{2},\alpha _{3})$
representing a set of parameters to be specified. In the following we use
this definition for calculating the physical components of the
energy-momentum tensor, $T^{\mu \lambda (11)}(x;\alpha )$, in different
situations.

\section{TFD stress-energy tensor and temperature}

As a basic result let us first calculate the temperature effect for the
electromagnetic field using this TFD approach. For such a proposal we assume
that $\alpha _{0}=i\beta =i/T$ and $\alpha _{1}=\alpha _{2}=\alpha _{3}=0\ $
in Eq.(\ref{genbog}). In this case we have 
\begin{eqnarray*}
\overline{G}^{(11)}(x-x^{\prime };\alpha ) &=&\frac{1}{(2\pi )^{4}}\int
d^{4}k\,e^{ik(x-x^{\prime })}\overline{G}^{(11)}(k;\alpha ) \\
&=&\frac{1}{(2\pi )^{4}}\int d^{4}k\,e^{ik(x-x^{\prime })}\sum_{j=1}^{\infty
}e^{\beta jk_{0}}[G_{0}(k)+\widetilde{G}_{0}(k)] \\
&=&2\sum_{j=1}^{\infty }G_{0}(x-x^{\prime }-i\beta jn),
\end{eqnarray*}
where $n^{\mu }=(1,0,0,0).$ Using this result in Eq.(\ref{emtenso1}) we find 
\begin{eqnarray}
T^{\mu \lambda (11)}(x;\alpha ) &=&-i\{\Gamma ^{\mu \lambda
}(x,x^{\prime }) \overline{G}^{(11)}(x-x^{\prime };\alpha
)\}|_{x\rightarrow x^{\prime }}, \nonumber \\
&=&-i\{4(\partial ^{\mu }\partial ^{\prime \lambda
}-\frac{1}{4}g^{\mu \nu }\partial ^{\rho }\partial _{\rho
}^{\prime })\sum_{j=1}^{\infty }G_{0}(x-x^{\prime }-i\beta
jn)\}|_{x\rightarrow x^{\prime }},  \nonumber \\
&=&-\frac{2}{\pi ^{2}}\sum_{j=1}^{\infty }\frac{g^{\mu
\lambda }-n^{\mu }n^{\lambda }}{(j\beta )^{4}}  \nonumber
\\ &=&-\frac{1}{45}\pi ^{2}T^{4}(g^{\mu \lambda }-n^{\mu
}n^{\lambda }). \label{plu112} \end{eqnarray}
\qquad\ 

Therefore, the energy density of the photon gas is the well
known result $ E(T)=T^{00(11)}(x;\alpha )$ which leads to
the relation for the black-body radiation, i.e.
\begin{equation}
E(T)=\frac{1}{15}\pi ^{2}T^{4}.  \label{black12}
\end{equation}
We find that the condensation effect introduced via the Bogoliubov
transformation is equivalent, in the imaginary time formalism, to the
displaced images in time by $ij\beta $ giving rise to a cut off in time axis 
\cite{brown,plu1}. In the next section we will see as this condensation
produces a cutoff in a space axes.

\section{ Compactification in space-time and the Casimir
effect}

In this section we use the formalism developed in Section 2 to derive the
Casimir effect at zero and non-zero temperature. We proceed with the same
prescription for the energy-momentum tensor but with a proper definition of
the parameter $\alpha $.

\subsection{Casimir effect at zero temperature}

In the last section we have derived the temperature effect for the
electromagnetic field with a proper choice for \ the parameters $\alpha _{l}$
which leads to a Green
 function $\overline{G}^{(11)}(x-x^{\prime };\alpha _{l})$ written in terms
of a modified free Green
 function, $G_{0}(x-x^{\prime }-i\beta jn).$ Let us now assume that $\alpha
_{0}=\alpha _{1}=\alpha _{2}=0$ and $\alpha _{3},$ is a real parameter. In
this case we have 
\begin{eqnarray}
\overline{G}^{(11)}(x-x^{\prime };\alpha ) &=&\frac{1}{(2\pi )^{4}}\int
d^{4}k\,e^{ik(x-x^{\prime })}\overline{G}^{(11)}(k;\alpha )  \nonumber \\
&=&\frac{1}{(2\pi )^{4}}\int d^{4}k\,e^{ik(x-x^{\prime })}\sum_{l=1}^{\infty
}e^{ilk_{3}\alpha _{3}}[G_{0}(k)+\widetilde{G}_{0}(k)]  \nonumber \\
&=&2\sum_{l=1}^{\infty }G_{0}(x-x^{\prime }-\alpha _{3}lz).  \label{direc11}
\end{eqnarray}
Observe from this expression that if we define $\alpha _{3}=2d$, the sum
over $l$ defines the non trivial part of the Green function
that leads to  the Dirichlet   boundary
condition for the electromagnetic field, considering two
conducting  parallel plates, one at $x_{3}=0$ and the other
at $ x_{3}=d$, see Ref.\cite{brown,plu1}. Here the sum is,
equivalently interpreted as being, over half of even images
of a photon propagating between two parallel plates (the
factor 2 in the sum will take into account   the other
half of the even images when $x\rightarrow x^{\prime}$).
This result was first derived by Brown and Macley, who
showed that the contributions due to odd images add to
zero\cite{brown,plu1}. In our approach this fact is
obtained explicitly without reference to images, and the
cancellation of the equivalent odd images is a result of
the tilde-propagator contribution. Then the nature of the
boundary conditions over the electromagnetic field,
associated Green's function, is then the prescription to 
define the physical content of the parameters $\alpha$.

Using Eq.(\ref{direc11})\ in Eq.(\ref{emtenso1}), with $\alpha _{3}=2d$, we
find for the energy-momentum tensor 
\begin{eqnarray}
T^{\mu \nu (11)}(x;d) &=&-i\{\Gamma ^{\mu \nu }(x,x^{\prime
})\overline{G} ^{(11)}(x-x^{\prime };\alpha
)\}|_{x\rightarrow x^{\prime }},  \nonumber \\
&=&-i\{2(\partial ^{\mu }\partial ^{\prime \nu
}-\frac{1}{4}g^{\mu \nu }\partial ^{\rho }\partial _{\rho
}^{\prime })2\sum_{l=1}^{\infty }G_{0}(x-x^{\prime
}-2dlz)\}|_{x\rightarrow x^{\prime }},  \nonumber \\
&=&-\frac{2}{\pi ^{2}}\sum_{l=1}^{\infty }\frac{g^{\mu \nu
}+4z^{\mu }z^{\nu }}{(2ld)^{4}}  \nonumber \\
&=&-\frac{1}{2^{3}\pi ^{2}d^{4}}(g^{\mu \nu }+4z^{\mu
}z^{\nu })\sum_{l=1}^{\infty }\frac{1}{l^{4}}  \nonumber \\
&=&-\frac{\pi ^{2}}{180d^{4}}(\frac{1}{4}g^{\mu \nu
}+z^{\mu }z^{\nu }).
\label{plu113}
\end{eqnarray}

In particular the component of the Casimir energy, $T^{00(11)}(x;\alpha
)=E(d)$, is then given as 
\begin{equation}
E(d)=-\frac{\pi ^{2}}{720d^{4}}.  \label{black13}
\end{equation}

\subsection{Casimir effect at non-zero temperature}

Consider the case $\alpha _{0}=i\beta ,\alpha _{1}=\alpha _{2}=0,$ $\alpha
_{3}=2d,$ $n^{\mu }=(1,0,0,0)$ and $z^{\mu }=(0,0,0,n_{3}),$ In this case we
have 
\begin{eqnarray*}
\overline{G}^{(11)}(x-x^{\prime };\alpha ) &=&\frac{1}{(2\pi )^{4}}\int
d^{4}k\,e^{ik(x-x^{\prime })}\overline{G}^{(11)}(k;\alpha ) \\
&=&\frac{1}{(2\pi )^{4}}\int d^{4}k\,e^{ik(x-x^{\prime
})}\sum_{j,l=1}^{\infty }e^{\beta
jk_{0}+ilk_{3}(2d)}[G_{0}(k)+\widetilde{G} _{0}(k)] \\
&=&2\sum_{j,l=1}^{\infty }G_{0}(x-x^{\prime }-i\beta jn-2dlz).
\end{eqnarray*}
Using this result in Eq.(\ref{emtenso1}) we find that the energy-momentum
tensor is 
\begin{eqnarray}
T^{\mu \nu (11)}(x;d,\beta ) &=&-i\{\Gamma ^{\mu \nu
}(x,x^{\prime }) \overline{G}^{(11)}(x-x^{\prime };\alpha
)\}|_{x\rightarrow x^{\prime }} \nonumber \\
&=&-i\{2(\partial ^{\mu }\partial ^{\prime \nu }-\frac{1}{4}g^{\mu \nu
}\partial ^{\rho }\partial _{\rho }^{\prime })  \nonumber \\
&&\times 2\sum_{j,l=1}^{\infty }G_{0}(x-x^{\prime }-i\beta
jn-2dlz)\}|_{x\rightarrow x^{\prime }}.  \label{plu114}
\end{eqnarray}
Notice that if we take in the sum given in Eq.(\ref{plu114}) $l=0$ or $j=0,$
we recover the terms of the black body radiation, Eq.(\ref{black12}), and
Casimir effect at zero temperature, Eq.(\ref{black13}). Thus in this case
the energy-momentum tensor is 
\begin{eqnarray}
T^{\mu \nu (11)}(x;d,\beta )&=&-\frac{4}{\pi
^{2}}\sum_{j,l=0^{\prime }}^{\infty }\{\frac{g^{\mu \nu
}}{[(2ld)^{2}+(j\beta )^{2}]^{2}} \nonumber \\
&&+\frac{ 4(2ld)^{2}z^{\mu
}z^{\nu}-(j\beta )^{2}n^{\mu }n^{\nu }}{
[(2ld)^{2}+(j\beta )^{2}]^{2}}\},
\end{eqnarray}
where the notation $j,l=0^{\prime }$ is used to emphasize
that the term for $ l=j=0$ is not included in the sum
(actually this is a divergent term which was subtracted in
Eq.(\ref{casad1})).

Following Brown and Maclay\cite{brown}
, we define $\xi =d/\beta $ , \[
f(\xi )=-\frac{1}{4\pi ^{2}}\sum_{j,l=0^{\prime }}^{\infty
}\frac{(2\xi )^{4} }{[(2l\xi )^{2}+(j)^{2}]^{2}},
\]
and 
\begin{eqnarray*}
s(\xi ) &=&-\frac{d}{d\xi }f(\xi ) \\
&=&\frac{2^{4}}{\pi ^{2}}\sum_{j,l=0^{\prime }}^{\infty
}\frac{\xi ^{3}j^{2} }{[(2l\xi )^{2}+(j)^{2}]^{3}}
\end{eqnarray*}
resulting in 
\[
T^{\mu \nu (11)}(x;d,\beta )=\frac{1}{d^{4}}f(\xi
)(g^{\mu \nu }+4z^{\mu }z^{\nu})+\frac{1}{\beta
d^{3}}(n^{\mu }n^{\nu }+z^{\mu }z^{\nu })s(\xi ).
\]
In particular the component $T^{00(11)}(x;d,\beta )=E(d,\beta )$ gives rise
to the energy density 
\[
E(d,\beta )=\frac{1}{d^{4}}[f(\xi )+\xi s(\xi )].
\]
Here $f(\xi )$ describes the Helmholtz free-energy density for photons and $
s(\xi )$ is the entropy density.

\section{Casimir-Boyer model}

In the last section we applied the generalized Bogoliubov transformation to
treat the Casimir effect, such that the Green
function fulfilled the  Dirichlet( Neumann) boundary condition for two
conducting (permeable) parallel plates, one at $x_{3}=0$ and the other at $
x_{3}=d.$ In this section we consider the Casimir-Boyer model\cite{boyer2},
corresponding to a mixed situation of plates in which at $x_{3}=0$ we have a
conducting plate (Dirichlet boundary conditions) and at
$x_{3}=d,$ a permeable plate (Neumann boundary
conditions). In order to have a Green function satisfying
these conditions, we consider $\alpha _{0}=i\beta ,\alpha
_{1}=\alpha _{2}=0,$ $\alpha _{3}=2d+\pi /k_{3},$ $n^{\mu
}=(1,0,0,0) $ and $z^{\mu }=(0,0,0,1).$ In this case we
have
\begin{eqnarray*}
\overline{G}^{(11)}(x-x^{\prime
};\alpha ) &=&\frac{1}{(2\pi )^{4}}\int
d^{4}k\,e^{ik(x-x^{\prime })}\overline{G}^{(11)}(k;\alpha )
\\ &=&\frac{1}{(2\pi )^{4}}\int d^{4}k\,e^{ik(x-x^{\prime
})}\sum_{j,l=1}^{\infty }(-1)^{l}e^{\beta
jk_{0}+ilk_{3}(2d)}[G_{0}(k)+ \widetilde{G}_{0}(k)] \\
&=&2\sum_{j,l=1}^{\infty }(-1)^{l}G_{0}(x-x^{\prime
}-i\beta jn-2dlz).
\end{eqnarray*}
Using this result in
Eq.(\ref{emtenso1}) we find the energy-momentum tensor to
be \begin{eqnarray}
T^{\mu \nu (11)}(x;d,\beta ) &=&-i\{\Gamma ^{\mu \nu }(x,x^{\prime })
\overline{G}^{(11)}(x-x^{\prime };\alpha )\}|_{x\rightarrow x^{\prime }} 
\nonumber \\
&=&-i\{2(\partial ^{\mu }\partial ^{\prime \nu }-\frac{1}{4}g^{\mu \nu
}\partial ^{\rho }\partial _{\rho }^{\prime })  \nonumber \\
&&\times 2\sum_{j,l=1}^{\infty }(-1)^{l}G_{0}(x-x^{\prime }-i\beta
jn-2dlz)\}|_{x\rightarrow x^{\prime }}.  \label{direc115}
\end{eqnarray}

As before, if we take the sums starting from $l,j=0$, we include the Casimir
effect at zero temperature and the black body radiation. Hence, carrying out
the calculations in Eq.(\ref{direc115}), we find that 
\begin{eqnarray}
T^{\mu \nu (11)}(x;d,\beta ) &=&-\frac{4}{\pi
^{2}}\sum_{j,l=0^{\prime }}^{\infty }(-1)^{l}\{\frac{g^{\mu
\nu }}{[(2ld)^{2}+(j\beta )^{2}]^{2}} \nonumber \\
&&+\frac{4(2ld)^{2}z^{\mu }z^{\nu }-(j\beta
)^{2}n^{\mu }n^{\nu }}{ [(2ld)^{2}+(j\beta
)^{2}]^{2}}\}.  \label{direct117}
\end{eqnarray}
Observe that for the term $l=0,$ the component $T^{00(11)}(x;d,\beta )$ is
the black body radiation tem given in Eq.(\ref{black12}), and for $j=0$ we
have $T^{00(11)}(x;d,\beta )=E(d)$
\[
E(d)=\frac{7}{8}\frac{\pi ^{2}}{720d^{4}},
\]
which is the Casimir energy for the Casimir-Boyer model\cite
{boyer2,boy2,boy3}. Notice that this energy corresponds to an attractive
force which is $-7/8$ of the Casimir energy for plates of the same material.

Using $\xi =d/\beta $ , we  introduce
\[
\widehat{f}(\xi )=-\frac{1}{4\pi ^{2}}\sum_{j,l=0^{\prime }}^{\infty
}(-1)^{l}\frac{(2\xi )^{4}}{[(2l\xi )^{2}+(j)^{2}]^{2}}, 
\]
and 
\begin{eqnarray*}
\widehat{s}(\xi ) &=&-\frac{d}{d\xi }f(\xi ) \\
&=&\frac{2^{4}}{\pi ^{2}}\sum_{j,l=0^{\prime }}^{\infty }(-1)^{l}\frac{\xi
^{3}j^{2}}{[(2l\xi )^{2}+(j)^{2}]^{3}}
\end{eqnarray*}
resulting in 
\[
T^{\mu \nu (11)}(x;d,\beta
)=\frac{1}{d^{4}}\widehat{f}(\xi )(g^{\mu \nu
}+4z^{\mu }z^{\nu })+\frac{1}{\beta d^{3}}(n^{\mu
}n^{\nu }+z^{\mu }z^{\nu })s(\xi ). \]
In particular, the component $T^{00(11)}(x;d,\beta
)=E(d,\beta )$ gives rise to the energy density
\[
E(d,\beta )=\frac{1}{d^{4}}[\widehat{f}(\xi )+\xi \widehat{s}(\xi )]. 
\]
Here  $\widehat{f}(\xi )$ describes the Helmholtz free-energy density
for photons and $s(\xi )$ is the entropy density.

\section{Concluding remarks}

Summarising, in this work a generalization for the thermofield dynamics
(TDF) formalism is presented, via an analytic continuation
of the usual Bogoliubov transformation, in order to
describe a field in a confined region in space. We apply
the method to calculate the energy-momentum tensor of the
electromagnetic field in different situations associated
with the Casimir effect, such that in each case some
peculiar aspect of the approach are emphasized.

Our aim with the applications was  to
demonstrate that our calculational method simplifies the
study of the Casimir effect considerably, using the notion
of covariance throughout the calculations. Furthermore the
case that gives rise to repulsive force, which is of some
interest in the most recent
literature\cite{kenne,boy2,boy3},  is dealt with detail,
resulting in the following new results: a) the explicit
expression for the energy-momentum tensor; b) the explicit
calculation of the expressions for the Helmholtz free
energy, the internal energy and entropy.

In this TFD generalization, the Casimir effect is interpreted as a process
of condensation of the electromagnetic field. In the case of the Casimir
effect at zero temperature, the quasi-particles are described (for an
arbitrary mode) by
\begin{eqnarray*}
a(\alpha ) &=&u(\alpha )a-v(\alpha )\widetilde{a}^{\dagger } \\
\widetilde{a}^{\dagger }(\alpha ) &=&u(\alpha )\widetilde{a}^{\dagger
}-v(\alpha )a,
\end{eqnarray*}
with $a(\alpha )$ and $\widetilde{a}$ fulfilling the
canonical algebra of the creation and destruction
operators, that is $[a(\alpha ),a^{\dagger }(\alpha )]=
[\widetilde{a}(\alpha ),\widetilde{a}^{\dagger }(\alpha
)]=1.$ From these operators a  vacuum state $|0(\alpha
)\rangle $ can be defined,  such that $a(\alpha
)|0(\alpha )\rangle =0$. Therefore, regarding the
operators $a$ and $a^{\dagger },$  the state $|0(\alpha
)\rangle $ describes a condensate, as is the case for
the temperature in the usual TFD. This provides an unusual
insight into the role of vacuum in developing the Casimir
force. Thus not only the notion of vacuum but also its
structure (seen as a condensate) are crucial in producing
the Casimir effect. In a broader sense, this notion of
condensate is a central aspect thoroughout  the paper,
which can not be  derived in the context of the Matsubara's
formalism.

 In Ref.\cite{jorge2}, using a
modification of the Matsubara approach to treat spontaneous
symmetry breaking in compactified $\lambda\phi^4$ theory
and superconducting transition temperature in thin films,
it is shown  how to describe
a general space confinement of a field, not necessarily in
the ground state  as is the case of the Casimir effect.
 However, our contention is that a more
refined understanding of that modified Matsubara method
would be interesting, considering theoretical and practical
applications. This aspect has been achieved here by using
a generalization of TFD, which can  be  used as well
for the  systems studied in \cite{jorge2}.

The Matsubara formalism has also
been applied in the derivation of the so-called Lifshitz
formula, describing the Casimir force in real media
$\epsilon({\bf x},\beta)$ (not $ \epsilon_0=1$)
\cite{mostep2}. Recently, this formula has been
successfully used with the dielectric permittivity, as
is given by the Drude model function, to describe  the
Casimir force at nonzero temperature between real metals
\cite{mostep02}. Thereof, it would be interesting to
analyze the connection of our method and the Matsubara
approach more closely. This can be carried out by using
functional methods in TFD \cite{seme}. In our case,
considering the scalar field, we can start with the
following definition for the generating funcional
\begin{equation} {\cal Z} ^{ab}=N\exp\{ \frac{-i}{2}\int
(J \ \  -J)G(x-x',\alpha)\left( \begin{array}{c}
J\  \\
-J\
\end{array}
\right)\} dxdy,
\end{equation}
where the matrix $G(x-x',\alpha)$ is given in
Eq.(\ref{eq24}).
From ${\cal Z} ^{ab}$ an effective action, say $W$, can
then  be introduced by $W^{ab}=-i\ln {\cal Z}^{ab}$. Taking
$a=b=1$, we recover the Matsubara method, and in particular
the results given in Eq.(3.10) of Ref.\cite{mostep2},
 for the the effective action under the
zeta-functional regularization, the starting point to
derive the aforementioned Lifshitz formula. These aspects
regarding the use of generalized Bogoliubov transformation
associated with Casimir effect for real (not only ideal)
media  will be discussed in more detail elsewhere.

Ending these remarks, it is worthy to add that as this
method based on the Bogoliubov transformation is
independent of the type of field involved, it
should be useful to analyze the  Casimir effect in
the case of non-abelian gauge fields like quantum
chromodynamics. In this case the Casimir effect affects the
formation of the quark-gluon plasma, and as a consequence,
the phase transition from a confined to a deconfined state.

\begin{description}
\item  {\bf Acknowledgments: } The authors thank Professor M. Revzen for the
stimulating discussions and for his interest in this work.
We also thank the Referee for the detailed commentary and
the useful suggestions. One of us (AES) thanks A. P. C.
Malbouisson, J. M. C. Malbouisson for the interesting
discussions. This work was supported by CNPQ of Brazil and
NSERC of Canada. \end{description}

\end{document}